\documentclass[12pt]{iopart}
\usepackage{epsfig}
\begin{document}

\title[]{Test results of a prototype designed to detect horizontal cosmic ray flux}

\author{M. Iori, A. Sergi and D. Fargion  
}

\address{University of Rome ``La Sapienza'' , Rome, Italy}


\ead{\mailto{maurizio.iori@roma1.infn.it}, \mailto{antonino.sergi@roma1.infn.it}, \mailto{daniele.fargion@roma1.infn.it}}

\begin{abstract}
In this paper we report test results from a prototype
designed to detect muons from horizontal air shower at large zenith angle,
 $86^{0}<\Theta<93^{0}$.
To detect horizontal tracks and their directions we select them
according the muon vertical equivalent charge and we measure the
time of flight with a time resolution of 800 ps. Several
measurements are collected at different zenith angles. The
background studies performed with two modules show that the main
source is due to tracks crossing the module at the same time.
The upper limit of background flux for a single twin module 
is estimated to be
$10^{-9} ~cm^{-2}$ $s^{-1}$ $sr^{-1} (90\%CL)$.
We estimated the size of the surface array necessary to detect the
shower flux of the order of $10^{-9}~ cm^{-2}$ $yr^{-1}$ $sr^{-1}$ if originated
by Tau Air-Showers secondaries of GZK neutrino Tau below the
horizons.
\end{abstract}




\section{Introduction}
\begin{figure}
\includegraphics*[width=\textwidth]{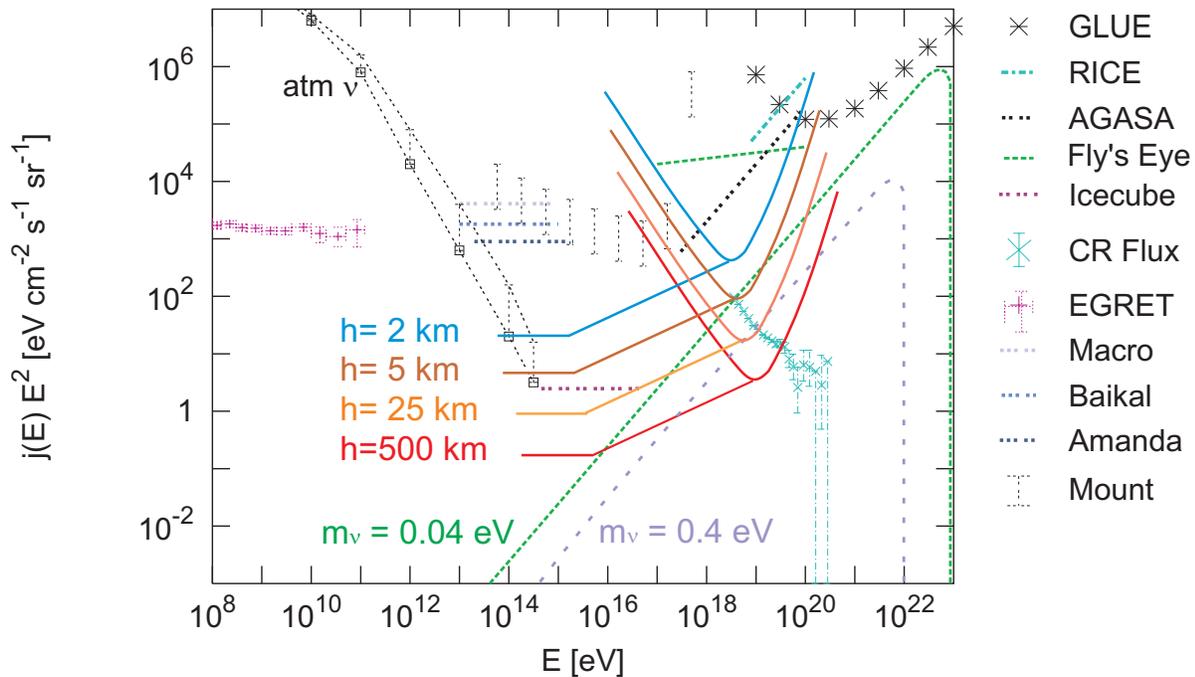}
\caption{\label{fig:scheme} UPTAUs (lower bound on the center)
and HORTAUs (right parabolic curves) sensibility at different
observer heights h ($2,5,25,500 km $) for a present neutrino flux
estimate in Z-Shower (Z-Burst) model scenario \cite{farg0},
\cite{Kalashev:2002kx}, for light ($0.4-0.04$ eV) neutrino
masses $m_{\nu}$; two corresponding density contrast has been
assumed \cite{farg3}; the lower parabolic bound thresholds are at
different operation height, in Horizontal (Crown) Detector
\cite{farg3} facing toward most distant horizons edge; we are
assuming a duration of data records of a decade comparable to the
BATSE record data (a decade). The parabolic bounds on the EeV
energy range in the right sides are nearly un-screened by the
Earth opacity while the corresponding UPTAUs bounds in the
center below suffer both of Earth opacity as well as of a
consequent shorter Tau interaction lenght in Earth Crust, that
has been taken into account. All present estimate are found for
the Earth crust made of water; for the rock present bounds are
lowered and sharper by nearly a factor $2.5$ \cite{farg3}. }
\end{figure}

One of the interesting subject in Astroparticle Physics is the
understanding of the origin and the composition of the energy
spectrum of Ultra High Energy Cosmic rays (UHE) that exceed
 $ 4 \cdot 10^{19}$ eV. These UHECR are just near the so called GZK cut-off
opacity due to cosmic black body radiation. These UHECR should be
also source of secondary neutrinos by photopion production made by
GZK cut-off. Many experiments detect neutrinos by looking for
showers or tracks from charged leptons produced by charged current
interaction of neutrinos of $10^{14}-10^{16}$ eV. At present time
there are several measurements of cosmic ray flux made with
several techniques \cite{agasa,ice,fly,auger,sudan}.
 Next generation experiment above $10^{19}$ eV
is sensitive to neutrinos typically from the near horizontal
air-showers. These projects include the Southern site of Pierre
Auger Observatory able to detect by photo-luminescent lights
along the horizontal showers ($\Theta < 60^{0}$). Recently several
authors have pointed the interest to the horizontal showers
produced by the interaction of ultra high energy (UHE) neutrinos
on the Earth crust \cite{farg0,farg1,farg2,Feng}. A measurement of
horizontal showers provides a test of the GZK model as an indirect
constraints of the ultra-high energy neutrino fluxes
\cite{semikoz}. Fig. \ref{fig:scheme} shows the neutrino fluxes
per flavour as predicted by Top-Down model with p=1 and
$m_{X}=2\cdot10^{13}$GeV/$c^{2}$ and compared to future
experimental sensitivity, \cite{semikoz}.
In this paper we present a prototype module designed to detect
horizontal cosmic ray flux and we show the test results obtained
in laboratory for a single module as for two modules. We discuss
also a possible engineering array to reach a sensitivity of the order
of $10^{-9}$ $cm^{-2}$ $yr^{-1}$ $sr^{-1}$.
\section{A unit of the detector}
\begin{figure}[h!tbp]
\vspace{7.5cm}
\includegraphics{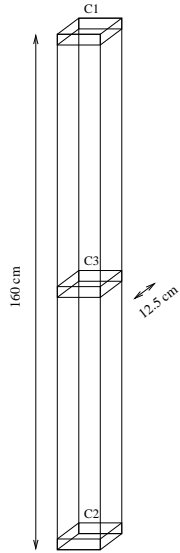}
\caption{\label{fig:scheme2} Prototype scheme}
\end{figure}

The basic idea is to measure the flux of particles from a shower produced
horizontally
in the atmosphere by the interaction of ultra high energy (UHE) neutrinos
on the Earth's crust.
The air shower is a complex chain of interactions resulting in a large
number of particles.
The predicted transverse length of the shower at ground level
should be about $1000-2000~ m$. The best solution to detect muons produced
by horizontal air shower consists in assembling several detectors pointing
the horizon from the top of a mountain in an large surface array.
A good time correlation of the signals of about 20 ns
provides the time structure of muonic showers and a time resolution
along the shower propagation of about 1 ns provides the background rejection
as discussed in next section.
The array will be discussed in section \ref{sec:array}.
The elementary module, called ``tower'' throughout this paper,
tested is composed by three tiles of Bicron
scintillator of 12.5 $\times$ 12.5 $cm^2$ and $2~ cm$ of thickness
put parallel in a frame shown in Fig.\ref{fig:scheme2}.
The three tiles in the paper are called C1, C2 and C3 and they are apart
between them $80~ cm$.
Two of them, C1 and C2, located at $x$ coordinate equal $0~ cm$ and
$x$ coordinate equal $160~ cm$
respectively, are
used to define the electronic coincidence; the third tile, located
between them, is used to study and to remove the background.
The distance of the centers of the tiles can
be varied from $2~ cm$ to $210~ cm$. The scintillators are read
by a Hamamatsu Photo-Multiplier coupled to the scintillator by an optical
connector.
The PM has a transit time spread of less than $0.8~ ns$ and a low-voltage power
supply. This choice requires a simple power supply system.
To measure horizontal cosmic ray flux we need know the time arrow, i.e. select
the muons from the horizon and remove particles from
other directions.
That requires the measurement of the time of flight between two tiles (C1,C2).
The results shown were performed with a setup where the distance between
C1 and C2 was $160~ cm$ and the data were taken at different zenith angles
and at ground level. One test was performed also with the C3 tile to
understand the backgroud.
\subsection{The background }
\label{sec:backgr}
The measurement of a cosmic particle passing through two tiles
distant $160~cm$ is affected by three kinds of background:
\begin{itemize}
\item a) a pair of different particles cross at the same time the
two tiles (C1,C2).
\item b) a particle crosses one tile and the noise of
second tile gives a coincidence
\item c) two particles from different mini-showers cross the first (C1) and 3rd (C2) tile at
different time.
\end{itemize}
The background a) can be removed by a good timing:
$\sigma (t_{1}-t_{2})\simeq 0$; the background
b) can be removed appling a cut on the charge deposited into the
scintillators;
The background c) must be evaluated and it cannot be removed.
That defines our response in term of signal$/$noise.
\subsection{Detector calibration and definition of an horizontal track}
\begin{figure}[tbp]
\begin{center}
\includegraphics[width=0.8 \textwidth, height=6 cm ]{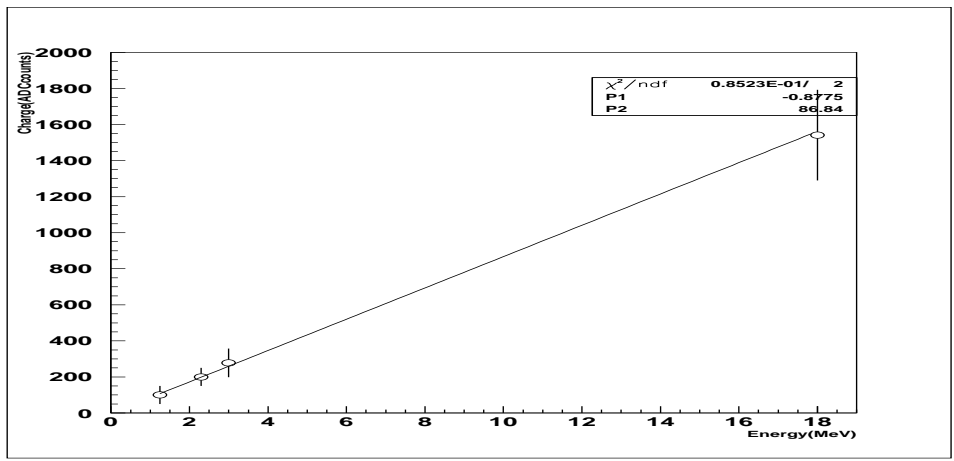}
\caption{\label{fig:calib} Charge in ADC counts as
function of energy lost by the particle}
\end{center}
\end{figure}

\begin{figure}
\begin{center}
\includegraphics[width=0.75 \textwidth, height= 4 cm]{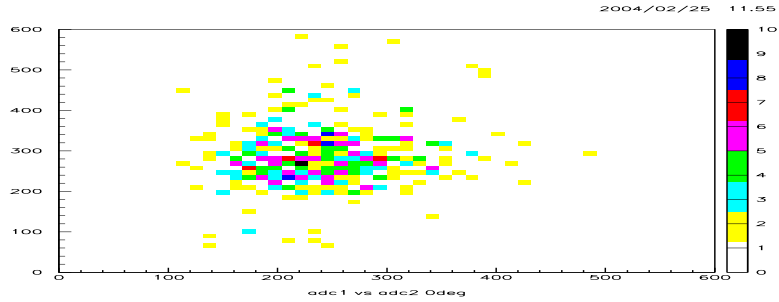}
\caption{\label{fig:adc1vs2} Charge deposited by a muon crossing the two tiles, C1 and C2 at 0$^{0}$ zenith angle.  }
\end{center}
\end{figure}

We calibrated the detector using $Sr^{90}$ and $Co^{60}$ radioactive
sources and the cosmic rays. For $Sr^{90}$ and $Co^{60}$ we used the
end-point of the spectrum, even for $Co^{60}$, because of the poor energy
resolution; for cosmic rays we assumed mimimum ionization particles
and changed the mean crossing length altering the angle of the
central tile with the others, from parallel ($3MeV$) to orthogonal ($18MeV$, 
corresponding to $12.5~cm$ path length in the scintillator).
Fig.\ref{fig:calib} shows the charge
deposited into the scintillator as function of the energy lost.
To isolate the particle crossing orthogonal the detector we
registered a sample of cosmic rays with the detector positioned
at zenith angle equal $0^{0}$ degrees. Analyzing the charge
deposited into the scintillator we found a set of cuts to define
a particle crossing the detector within a solid angle of
$4.8\cdot 10^{-3}sr$. The sample of events selected by these
cuts is still contaminated by a background of type a), described
above. The time of flight and the information of the energy
deposited in C2 will remove this contribution.
Fig.\ref{fig:adc1vs2} shows the charge deposited into tile C1 and
C2 by a vertical muon of energy greater than 1 $GeV$.
The poor energy resolution is due to the light collection method:
to enhance our time resolution we decided to remove any light
yield enhancement technique and take only the light that travels
on a straight line, from the interaction point to the photocathode;
more precisely, we observed that any further reflection is harmful
for time resolution, so we decided to loose the energy resolution
to make direct light collection dominant.
The charge of vertical equivalent muon (CVEM), as discussed before,
is used to define a set of cuts ($\pm$ 2.5 nC around the average
deposited charge) to select horizontal charged tracks. It also
let us remove events falling in regions of the scintillator
affected by a large time spread in light collection, giving us an
enhancement in time resolution; this has been proved by testing
the light collection time uniformity of a single tile, by
coincidence with a thin plastic scintillator of signals from a
$Sr^{90}$ source.

\subsection{The time resolution}
\begin{figure}
\begin{center}
\includegraphics[width=0.75 \textwidth, height= 4 cm]{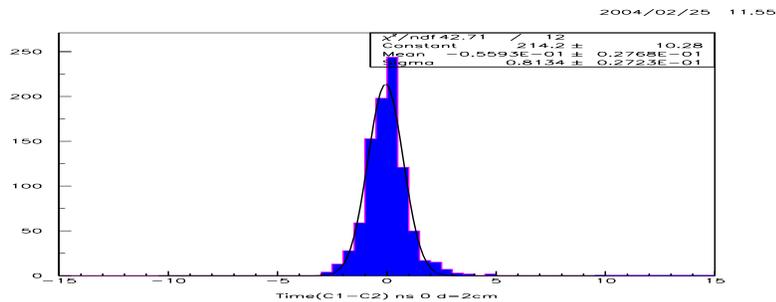}
\caption{\label{fig:tres} $\Delta t$ (ns) between C1 and C2 when they are $2~ cm$ distant.}
\end{center}
\end{figure}

\begin{figure}
\begin{center}
\includegraphics[width=0.75 \textwidth, height= 4 cm]{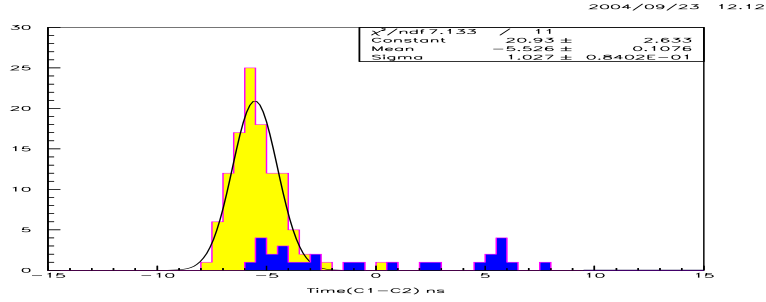}
\caption{\label{fig:horiz1} $\Delta t$ (ns) between C1 and C2
when their centers are $160~ cm$ distant; The yellow histogram is obtained
with the detector at zenith angle equal $0^{0}$; the blue
histogram at zenith angle equal $90^{0}$. The data correspond to
29 hr and 65 hr live time respectively. Results statistically
consistent are obtained if we reverse the counters (i.e C2 top,
C1 bottom). The events at $\Delta t\sim 0$ are produced by two
parallel tracks.
 }
\end{center}
\end{figure}

To evaluate the time resolution of the apparatus we put the centers of
two tiles
at a distance of $2~cm$. The distribution obtained, shown in
Fig.\ref{fig:tres}, has a $\sigma=0.85~ ns$. Then an event with
$\vert \Delta t\vert <2.5~ ns$ is not a single track passing
through C1 and C2.
Fig.\ref{fig:horiz1} shows the time response when the detector is
at zenith angle equal $0^{0}$ (yellow). Being the distance
between the centers of the two tiles $160~ cm$ the time of flight is $5.8 \pm
0.8 $ ns. The distribution with the detector oriented at zenith
equal $90^{0}$ is also plotted (blue). The peaks at $\vert t
\vert=5.8~ ns$ correspond to horizontal tracks crossing C1 then
C2 or viceversa. The events at $t\sim 0~ ns$ are generated by two
tracks that cross at the same instant the two tiles.
\section{Angular detection of muon flux by one tower and evaluation of the background}
The results of about 400 hours live time recorded at lowest of a
four floor building are summarized in Table \ref{tab:meas}. At
that place we are screened by most of the direct gamma shower
secondaries in open air.
\begin{table}
\begin{center}
\begin{tabular}{|c|c|c|c|c|c|}
\hline
 deg & Run time & C1 10$^{-3}$Hz& C2 10$^{-3}$Hz &$I_{\mu}$ $cm^{-2}s^{-1}sr^{-1}$& Fig \\[1ex]
\hline
 0 & 29h:48m & 6.44$\pm$0.05 & $<$0.002 & 7.9$\pm$0.9 $\times$ $10^{-3}$& \ref{fig:horiz1} \\ [1ex]
 86.5-93.5& 65h:26m$^{1}$ & 0.034$\pm$0.010 & 0.042 $\pm$0.010& 4.2$\pm$1.5 $\times$ $10^{-5}$& \ref{fig:horiz1} \\ [1ex]
 86.5-93.5& 125h:00m$^{2}$ & 0.025$\pm$0.01 &  & 3.2$\pm$1.5 $\times$ $10^{-5}$& \ref{fig:horiz2} \\ [1ex]
 86.5-93.5& 125h:00m$^{2}$ &  & 0.020 $\pm$0.01& 2.5$\pm$1.3 $\times$ $10^{-5}$& \ref{fig:horiz2} \\ [1ex]
 81.6-88.8& 111h:11m $^{2}$ & 0.045 $\pm$ 0.01&  &6.6$\pm$1.5 $\times$ $10^{-5}$ & \ref{fig:horiz3} \\ [1ex]
 81.6-88.8& 111h:11m $^{2}$ & &  0.015 $\pm$ 0.01 &2.2$\pm$0.9 $\times$ $10^{-5}$ & \ref{fig:horiz3} \\ [1ex]
 75.5-81.8& 111h:11m & 0.130 $\pm$ 0.002 & & 1.9$\pm$0.3 $\times$ $10^{-4}$& \ref{fig:horiz4} \\ [1ex]
 75.5-81.8& 111h:11m & &0.0025$\pm$0.0025& 3.7$\pm$3.7 $\times$ $10^{-6}$ &
 \ref{fig:horiz4} \\ [1ex]
 \hline
\end{tabular}
\caption{\label{tab:meas} Cosmic ray flux measured by one unit only;$^{1}$ the detector is
located at $80~ cm$ of floor level E-W direction;
$^2$ for these measurements
the setup is show in Fig. \ref{fig:setup} and described in the text}
\end{center}
\end{table}

\begin{figure}[htbp]
\begin{center}
\includegraphics[width=12.cm,totalheight=8.cm]{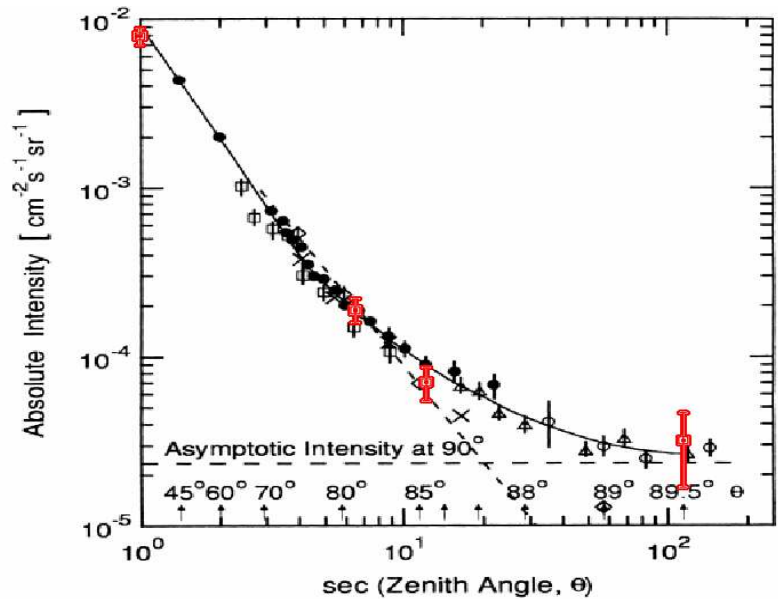}
\caption{\label{fig:graph1} Plot of the results (red points) from Table 1 in the interval 0$^{0}$,90$^{0}$ degrees; other
data are from \cite{cosmic}}
\end{center}
\end{figure}

In table \ref{tab:meas} is evaluated the intensity of muon as
function of zenith angle corrected by detector efficiency. The
last six raw give the results obtained pointing the detector to a
step of concrete as shown in Fig.7. The step drawn on right side
(South-West) of the layout permits to the muons at large zenith
angle (greater than 88 degrees) cross at least 50 m of concrete. On
the left side (North) muons cross the detector at angle smaller
than 85 degrees directly from the atmosphere. The measured
difference of the muon flux from North and from South corresponds
to about 100 m water equivalent. Fig.\ref{fig:horiz3} shows the
C1-C2 time difference (blue histogram) for zenith angle
88.8$^{0}$-81.6$^{0}$ . Yellow histogram is obtained with the
detector at zenith equal $0^{0}$. The measured difference of
factor 2 in flux is due to the change of acceptance.
Fig.\ref{fig:horiz4}a shows the results with the detector at
zenith angle equal to 75.5$^{0}$-81.8$^{0}$. At this angle the
cosmic rays hit before C1 tile than C2 tile hence the time
difference defined $\Delta t_{12}$ is negative. The plot in Fig.
\ref{fig:horiz4}b shows the ADC counts of the C3 tile located
between C1 and C2 versus the $\Delta t_{12}$. By this plot we
estimate an upper limit to background (type c, cfr. \ref{sec:backgr}) 
of order of $10^{-6}~Hz$ for each
module. The results of Table 1 are plotted in fig.\ref{fig:graph1}.

\begin{figure}[htbp]
\begin{center}
\includegraphics[width=12.cm,totalheight=5.cm]{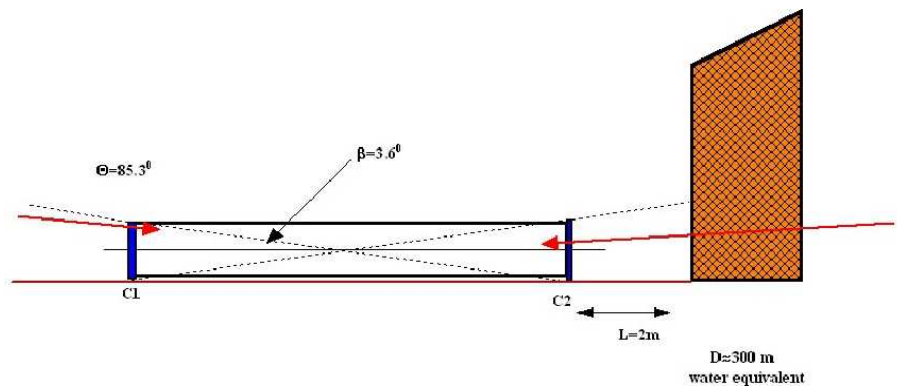}
\caption{\label{fig:setup} Setup to measure the effect on the muon flux
induced by a step of concrete }
\end{center}
\end{figure}

\begin{figure}
\begin{center}
\includegraphics[width=0.75 \textwidth, height= 4 cm]{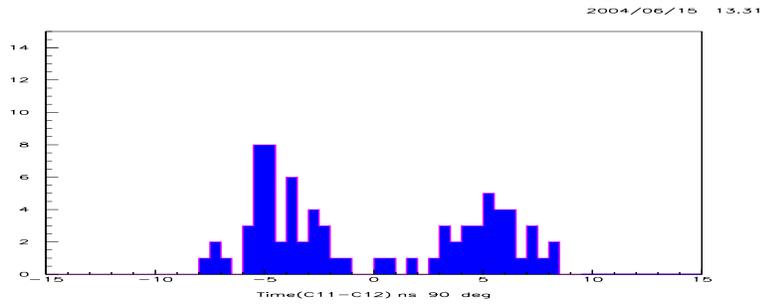}
\caption{\label{fig:horiz2} $\Delta t$(ns) between C1 and C2 when they are $160 cm$ distant. The data correspond to 125 hr live time. }
\end{center}
\end{figure}

\begin{figure}
\begin{center}
\includegraphics[width=0.75 \textwidth, height= 4 cm]{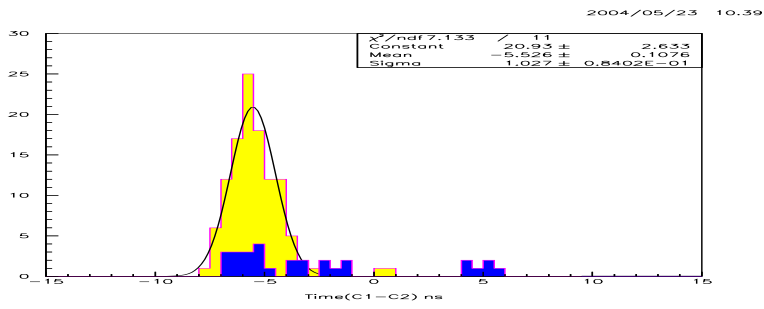}
\caption{\label{fig:horiz3} $\Delta t$(ns) between C1 and C2 when they are $160 cm$ distant;
The yellow histogram is obtained with the detector at zenith angle equal $0^{0}$; the blue
histogram at zenith angle equal $86^{0}$. The data correspond to 29 hr and 111 hr live time respectively. }
\end{center}
\end{figure}

\begin{figure}
\begin{center}
\includegraphics[width=0.75 \textwidth, height= 7 cm]{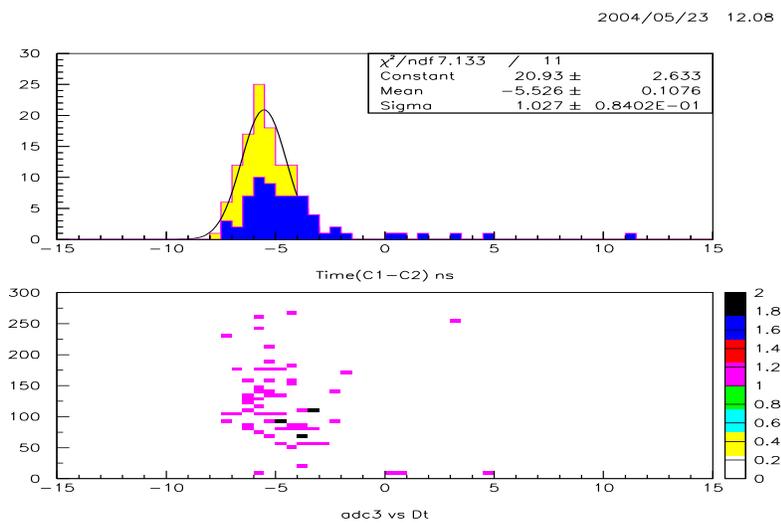}
\caption{\label{fig:horiz4}a) $\Delta t_{12}$(ns) between C1 and C2 when they are $160 cm$ distant;
The yellow histogram is obtained with the detector at zenith angle equal $0^{0}$; the blue histogram
at zenith angle equal $79^{0}$. The data correspond to 29 hr and 111 hr live time respectively;
b) ADC count of 3rd tile (C3) located in between C1-C2 versus $\Delta t_{12}$(ns). }
\end{center}
\end{figure}
\section{Test results with two towers}
To study the background and optimize the distance
of C1 and C2 tile we have inserted a second
module in the trigger gate.
\subsection{Setup at zenith angle $0^{0}$ }
We recorded the ADC counts
and the TDC signals from the two modules set at zenith $0^{0}$ and located
at distance $45~ cm$, $180~ cm$ and $250~ cm$. The rate of
two particles crossing vertically both modules,
distant $45~ cm$, in the trigger gate
was 3.7 $10^{-5}$ Hz. No events were found when the second module was
set vertically at distance $180~ cm$ and $250~ cm$.
These measurements tell us no background from vertical
is present when the tiles are $160~ cm$ distant or more.
\subsection{Setup at zenith angle $90^{0}$ to evaluate the sensitivity }
To complete these tests, due
to the fact our detector works horizontally ,we performed the same
measurement with the two modules at zenith angle $90^{0}$ and set parallel
at distance of $20~ cm$. In this setup we have measured a rate of
 1.5 $10^{-5}$ Hz of events when one track crosses one module and
\emph {only}
one tile of the second module was hit by another track at the same time.
The two modules were set at a distance of $300~ cm$ running for $1.7 \times 10^{6}$ s
(20 days) to estimate the background upper limit.
During this data taking we did not see events, then the sensitivity
is $10^{-9}$ $cm^{-2}s^{-1}sr^{-1}~(90\%CL)$, being our gate of about $250~ns$,
in agreement with muon bundles flux reported in Ref.\cite{decor}.
\section{The $t_{0}$ of the event and estimate of detector sensitivity}
\label{sec:array} 
\begin{figure}[htbp]
\begin{center}
\includegraphics[width=12.cm,totalheight=7.cm]{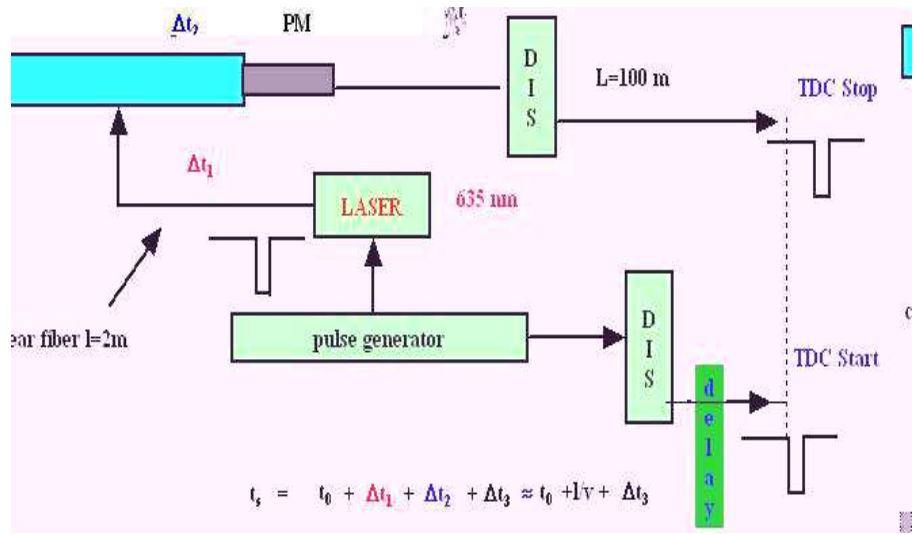}
\caption{\label{fig:calib3}Setup to measure the resolution on $t_{0}$ }
\end{center}
\end{figure}

The $t_{0}$ of the events recorded by each module can
be obtained using clear fibers connected to the tiles and
transporting a pulse of monochromatic laser light.
Fig.\ref{fig:calib3} shows the setup used to evaluate the time
resolution with two modules 300 m apart. It results to be less
that 10 ns. The measurement of the time of the signal transmitted
from the PM to the DAQ is evaluated recording the arrival time and
subtracting the time spent by the pulse to reach each tile:
the time resolution results to be less than 10 ns.
The flux detection limit can be estimated by
$$ \Phi(E)=\frac{N}{\Delta t A_{eff}\Delta \Omega \epsilon (E)}$$
where $N$ is the number of the events. $\Delta t$ the live time,
$A_{eff}$ the effective surface observed by the detector, $\Delta \Omega$
 the solid angle and $\epsilon (E)$ the neutrino
efficiency conversion. Neglecting the energy dependence of $A_{eff}$ and
assuming, for simplicity,
 $\epsilon (E)=1$ for Earth skimmed events, we obtain a detector sensitivity,
for frontal showers originated at about $100~Km$,
of 1.7 events/year with a flux of $10^{-11} cm^{-2} yr^{-1}sr^{-1}$
and a total detector live surface of $100~ m^{2}$. This is actually an
upper limit because of neglected geometrical and physical effects determined
by the actual \emph{in situ} configuration and has to be interpreted
as a fraction of the total flux, limited to Earth skimmed events; it
gives, for a $4\pi$ observer, a corresponding sensitivity of
about $10^{-9} cm^{-2} yr^{-1}sr^{-1}$, obtained by
rescaling for the ratio $4\pi/\Delta \Omega$, that estimates
a correction for the assumption of frontal showers. The assumption that
$100~ m^{2}$ of live surface is enough to achive this level of discrimination
is based on phenomenological studies \cite{farg1}.

\section{SUMMARY}
In conclusion the background measured for a single module is
sufficient for a reliable detection of muons from horizontal
showers if the distance of the tiles is between $160~ cm$ to
$200~ cm$. This setup permits also to have a detector sensitivity
of $10^{-9} cm^{-2} yr^{-1}sr^{-1}$, by an array with a live surface of $100~m^{2}$.
At this flux level the
Tau-Air Showers by GZK neutrinos might be already able to pollute
above the atmospheric muon fluxes \cite{farg3}. Therefore a
small size array at top of the mountains would be able to reach
neutrino astronomy at minimal guaranteed GZK fluence.

\section*{Acknowledgments}
M. I. thanks G. Boca for many discussions and the authors are indebted for invaluable technical support from F. Pulcinella.

\end{document}